\documentclass[sigconf,natbib=true, dvipsnames, svgnames]{acmart}

\AtBeginDocument{%
  \providecommand\BibTeX{{%
    \normalfont B\kern-0.5em{\scshape i\kern-0.25em b}\kern-0.8em\TeX}}}

\usepackage[inline]{enumitem}
\usepackage{pbox}
\usepackage{balance}
\usepackage{todonotes}
\newlist{inlinelist}{enumerate*}{1}
\setlist*[inlinelist,1]{label=\roman*),itemjoin={{; }},itemjoin*={{; and }}}
\usepackage{url}
\usepackage[normalem]{ulem}
\usepackage{multirow}
\usepackage{multicol}
\usepackage{booktabs}

\copyrightyear{2022}
\acmYear{2022}
\setcopyright{acmlicensed}\acmConference[SIGIR '22]{Proceedings of the 45th International ACM SIGIR Conference on Research and Development in Information Retrieval}{July 11--15, 2022}{Madrid, Spain}
\acmBooktitle{Proceedings of the 45th International ACM SIGIR Conference on Research and Development in Information Retrieval (SIGIR '22), July 11--15, 2022, Madrid, Spain}
\acmPrice{15.00}
\acmDOI{10.1145/3477495.3531857}
\acmISBN{978-1-4503-8732-3/22/07}

\begin{document}
\fancyhead{} 

\title{From Distillation to Hard Negative Sampling: Making Sparse Neural IR Models More Effective}




\author{Thibault Formal}
\email{thibault.formal@naverlabs.com}
\affiliation{%
  \institution{Naver Labs Europe}
  \city{Meylan}
  \country{France}
}
\affiliation{%
  \institution{Sorbonne Université, ISIR}
  \city{Paris}
  \country{France}
}

\author{Carlos Lassance}
\email{carlos.lassance@naverlabs.com}
\affiliation{%
  \institution{Naver Labs Europe}
  \city{Meylan}
  \country{France}
}

\author{Benjamin Piwowarski}
\email{benjamin@piwowarski.fr}
\affiliation{%
  \institution{Sorbonne Université, ISIR, CNRS}
  \city{Paris}
  \country{France}
}

\author{Stéphane Clinchant}
\email{stephane.clinchant@naverlabs.com}
\affiliation{%
  \institution{Naver Labs Europe}
  \city{Meylan}
  \country{France}
}

\begin{abstract}
Neural retrievers based on dense representations combined with Approximate Nearest Neighbors search have recently received a lot of attention, owing their success to distillation and/or better sampling of examples for training -- while still relying on the same backbone architecture. In the meantime, sparse representation learning fueled by traditional inverted indexing techniques has seen a growing interest, inheriting from desirable IR priors such as explicit lexical matching. While some architectural variants have been proposed, a lesser effort has been put in the training of such models. In this work, we build on SPLADE -- a sparse expansion-based retriever -- and show to which extent it is able to benefit from the same training improvements as dense models, by studying the effect of distillation, hard-negative mining as well as the Pre-trained Language Model initialization.
We furthermore study the link between effectiveness and efficiency, on in-domain and zero-shot settings, leading to state-of-the-art results in both scenarios for sufficiently expressive models.
\end{abstract}


\settopmatter{printacmref=true} 

\maketitle

\section{Introduction}

Traditional IR systems like BM25 have dominated search engines for decades~\cite{books/aw/Baeza-YatesR99}, relying on lexical matching and inverted indices to perform efficient retrieval. Since the release of large Pre-trained Language Models (PLM) like BERT~\cite{bert}, Information Retrieval has witnessed a radical paradigm shift towards contextualized semantic matching, where neural retrievers are able to fight the long-standing vocabulary mismatch problem.
In the first-stage ranking scenario, dense representations combined with Approximate Nearest Neighbors (ANN)  search have become the standard approach, owing their success to improved training pipelines.
While these models have demonstrated strong in-domain performance, their ability to generalize has recently been challenged on the recent zero-shot evaluation BEIR benchmark~\cite{beir_2021}, where their average effectiveness is lower than BM25 on a set of various IR-related tasks.

In the meantime, there has been a growing interest in going back to the ``lexical space'', by learning sparse representations than can be coupled with inverted indexing techniques. These approaches, which generally learn term weighting and/or expansion, benefit from desirable IR priors such as explicit lexical matching and decades of works on optimizing the efficiency of inverted indices.  
These have also shown good generalization capabilities -- w.r.t. either effectiveness~\cite{beir_2021} or IR behavior like exact match~\cite{formal2021match}.
While they mostly differ in their architectural design, a lesser effort has been put in the training of such models, making it unclear how they would be able to take advantage of the same improvements as dense architectures. In this work, we build on the SPLADE model~\cite{10.1145/3404835.3463098}, and study the effect of adopting the latest advances for training dense retrievers.
We provide an extensive experimental study -- by training models in various scenarios -- and illustrate the interplay between models capacity (as reflected by their sparsity) and performance. We show how improvements are additive, and how we are able to obtain state-of-the-art results for sufficiently expressive models.

\section{Related Works}
Replacing term-based approaches for candidate generation in search engine pipelines requires models and techniques that can cope with the high latency constraints of serving thousands of queries per second. Current approaches based on PLM either rely on dense representations combined with ANN search, or sparse representations that can benefit from the proven efficiency of inverted indices. 

\paragraph{\bf Dense Representation Learning} 
Dense retrieval has become the most common approach to apply PLM retrievers in IR or Question Answering. While dense models are generally similar -- e.g. relying on the \texttt{[CLS]} token -- various training strategies have recently been proposed to improve learned representations, ranging from distillation~\cite{hofstaetter2020_crossarchitecture_kd,lin-etal-2021-batch,Hofstaetter2021_tasb_dense_retrieval,santhanam2021colbertv2}, hard negative mining~\cite{xiong2021approximate,rocketqa_v1,ren-etal-2021-rocketqav2,DBLP:journals/corr/abs-2010-10469}, pre-training~\cite{DBLP:journals/corr/abs-2108-05540,izacard2021contriever} or any combination of the latter. These models -- including for instance DPR~\cite{karpukhin-etal-2020-dense}, ANCE~\cite{xiong2021approximate} or TAS-B~\cite{Hofstaetter2021_tasb_dense_retrieval} -- have recently been challenged on generalization tasks, as shown in the BEIR benchmark~\cite{beir_2021}. Among dense approaches, ColBERT~\cite{colbert} in contrast relies on a late-interaction mechanism, allowing to perform fine-grained term-level matching at the expense of higher indexing cost and latency.

\paragraph{\bf Sparse Representation Learning} In the meantime, sparse representations based on PLM have seen a growing interest, as they by design inherit from good properties of lexical models. These approaches generally rely on a term-importance component and/or a term-expansion mechanism. Various designs have been proposed in order to cope with one or both aspects. COIL~\cite{gao-etal-2021-coil} learns term-level dense representations to perform contextualized lexical match; uniCOIL~\cite{DBLP:journals/corr/abs-2106-14807} further simplifies the approach by learning a single weight per term, extending previous methods like DeepCT~\cite{10.1145/3397271.3401204} and DeepImpact~\cite{10.1145/3404835.3463030}. On the other hand, SPLADE~\cite{10.1145/3404835.3463098} directly learns high-dimensional sparse representations that are able to jointly perform expansion and re-weighting through the Masked Language Modeling head of the PLM and the help of sparse regularization.

\paragraph{\bf Motivation}
While dense approaches have benefited from various improved training strategies, it is unclear whether such improvements could also be observed for sparse models. We wonder if these improvements are \emph{additive}, in the sense that if a model is better than another in a ``normal'' training setting, would we still observe the same hierarchy in a distillation scenario? Answering such a question would allow to decouple architectures from training innovations when comparing neural retrievers. The recent ColBERTv2~\cite{santhanam2021colbertv2} demonstrates how ColBERT can leverage distillation and in-batch negatives to set-up the new state of the art on MS MARCO. In this paper, we follow a similar line: we build on SPLADE and extensively study the influence of various training strategies on model performance, for in-domain and zero-shot settings. 

\section{SPLADE and Methodology}\label{sec:3}


In this section, we first describe in details SPLADE~\cite{10.1145/3404835.3463098}, alongside various modifications that can be made in order to improve the model, from distillation and hard negative mining tricks, to the choice of the PLM.

\subsection{SPLADE}

\paragraph{\bf Model}
SPLADE is an expansion-based sparse retrieval model which predicts term importance in the BERT WordPiece vocabulary, relying on predictions from the Masked Language Modeling layer used at pre-training to implicitly perform term expansion.
For a query or document $t$, let $w_{i,j}$ denote the resulting importance of the $j$-th vocabulary token, for the input token $i$. Text representations are obtained by pooling such importance predictors over the input sequence, after a log-saturation effect. In SPLADE \cite{formal2021match}, \texttt{sum} pooling is used, but we found out experimentally that using \texttt{max} pooling, inspired by \cite{colbert}, led to major improvements over SPLADE (see Section~\ref{sec:4}). We thus consider by default the following formulation:
\begin{equation}\label{eq_splad_max}
w_j= \max_{i \in t} \log \left(1 + \text{ReLU}(w_{ij}) \right)
\end{equation}
and the ranking score $s(q,d)$ is given by dot product between $q$ and $d$ representations.

\paragraph{\bf Training} Given a query $q$, a positive document $d^+$, a negative document $d^-$ \emph{mined from BM25}, as well as additional in-batch negatives ${d_j^-}$ (i.e. documents from other queries in the given batch),
the model is trained by jointly optimizing a constrastive \texttt{InfoNCE} loss~\cite{Oord2018RepresentationLW} -- similarly to several prior works on learning first-stage rankers -- and the FLOPS regularization~\cite{paria2020minimizing} directly applied on representations, to obtain the desired sparsity for indices used at retrieval time:

\begin{equation}
\mathcal{L} = \mathcal{L}_{\texttt{InfoNCE},BM25} + \lambda_q \mathcal{L}^{q}_{\texttt{FLOPS}} + \lambda_d \mathcal{L}^{d}_{\texttt{FLOPS}}
\label{splade_loss}
\end{equation}

\subsection{Distillation, hard negative mining and PLM initialization}

In the following, we detail several training tricks previously introduced for dense models that can seamlessly be applied to SPLADE. All the extensions rely on a simple modification of Eq.~\ref{splade_loss}, either by modifying the ranking loss, the source of hard negatives, or a combination of both. We also discuss the initialization strategy, where models can further be improved by relying on PLM checkpoints that have been pre-trained on retrieval-oriented tasks.

\paragraph{\bf Distillation}

Distillation has been shown to greatly improve the effectiveness of various neural ranking models \cite{hofstaetter2020_crossarchitecture_kd,lin-etal-2021-batch,Hofstaetter2021_tasb_dense_retrieval,santhanam2021colbertv2}. We rely on the MarginMSE loss~\cite{hofstaetter2020_crossarchitecture_kd} -- MSE between the positive-negative margins of a cross-encoder teacher and the student, and train our model by optimizing\footnote{Cross-encoder teacher scores provided at \url{https://github.com/sebastian-hofstaetter/neural-ranking-kd}}:

\begin{equation}\label{mse_loss}
\mathcal{L} = \mathcal{L}_{\texttt{MarginMSE},BM25} + \lambda_q \mathcal{L}^{q}_{\texttt{FLOPS}} + \lambda_d \mathcal{L}^{d}_{\texttt{FLOPS}}
\end{equation} 

This distillation setting will be the default for all the scenarios introduced below.

\paragraph{\bf Mining hard negatives}
The standard setting using BM25 negatives is limited, and the benefit of using better negative sampling has been highlighted in several prior works~\cite{xiong2021approximate,rocketqa_v1,ren-etal-2021-rocketqav2,DBLP:journals/corr/abs-2010-10469}. Note that, by considering MarginMSE distillation as the basis for hard negative mining, we remove the need to resort on denoising, i.e. filtering noisy false negatives~\cite{rocketqa_v1,ren-etal-2021-rocketqav2}.

\paragraph{Self-mining} Following ANCE~\cite{xiong2021approximate}, which dynamically samples negatives from the model that is being trained, we propose to follow a simpler two-step strategy that has also been adopted in prior works~\cite{lin-etal-2021-batch}: \begin{itemize} \item (\texttt{step 1)} We initially train a SPLADE model, as well as a cross-encoder re-ranker, in the previously introduced distillation setting;
\item (\texttt{step 2)} We then generate triplets using SPLADE from \texttt{step 1}, and use the cross-encoder to generate the scores needed for MarginMSE, and go for another round of training.
\end{itemize}

This simply leads to a distillation strategy where mined pairs are supposed to be of better ``quality'' compared to BM25: \begin{equation}\label{eq_self}
\mathcal{L} = \mathcal{L}_{\texttt{MarginMSE},self} + \lambda_q \mathcal{L}^{q}_{\texttt{FLOPS}} + \lambda_d \mathcal{L}^{d}_{\texttt{FLOPS}}
\end{equation} 

\paragraph{Ensemble-mining}

While the self-mining scenario provides a better sampling strategy compared to BM25, it is rather limited as 
one could wonder if using various types of models to mine negatives for \texttt{step 2} could be beneficial. We rely on the recently released \texttt{msmarco-hard-negatives} dataset\footnote{\url{https://huggingface.co/datasets/sentence-transformers/msmarco-hard-negatives}}, available in the Sentence Transformers library \cite{reimers-2019-sentence-bert}, containing for each query: \begin{enumerate*}
    \item the top-$50$ hard negatives mined from BM25 and a set of $12$ various dense retrievers,
    \item the scores coming from a cross-encoder for each available $(q,d^+,d^-)$ in order to perform MarginMSE knowledge distillation.
\end{enumerate*}

\begin{equation}\label{eq_ensemble}
\mathcal{L} = \mathcal{L}_{\texttt{MarginMSE},ensemble} + \lambda_q \mathcal{L}^{q}_{\texttt{FLOPS}} + \lambda_d \mathcal{L}^{d}_{\texttt{FLOPS}}
\end{equation} 

\paragraph{\bf Pre-training}

NLP has recently borrowed ideas from contrastive learning techniques in Computer Vision, with the goal of learning high-quality sentence or document representations \emph{without annotation}~\cite{wu2020clear,DBLP:journals/corr/abs-2108-05540,izacard2021contriever,DBLP:journals/corr/abs-2112-07708,wang2021tsdae}. The general idea consists in designing pre-training tasks, that are better suited for subsequently training neural retrievers. Most approaches in IR are very similar to CoCondenser~\cite{DBLP:journals/corr/abs-2108-05540}, which constrastively learns the embedding space from spans of documents (two spans from the same document are considered as positives, the other documents in the batch as negatives). We thus simply consider to use such pre-trained checkpoint to initialize SPLADE, in the scenarios described above. Note that while CoCondenser is not directly optimized for sparse retrieval, its learned embedding space might contain more informative knowledge for retrieval compared to models pre-trained with Masked Language Modeling.


\section{Experiments and evaluation}\label{sec:4}

We conduct an extensive experimental study, by training and evaluating several models, for each scenario introduced in Section~\ref{sec:3}. As we provide improvements over SPLADE~\cite{10.1145/3404835.3463098}, we refer to all our strategies under the same alias, coined \texttt{SPLADE++}.

\subsection{Training and evaluation}

\paragraph{\bf Training}

All the models are trained on the MS MARCO passage ranking dataset\footnote{\url{https://github.com/microsoft/MSMARCO-Passage-Ranking}}, which contains $8.8M$ passages and $500k$ training queries with shallow annotations. We follow the same training and evaluation workflow described in \cite{10.1145/3404835.3463098}. Especially, we use the same hyperparameters, as well as the validation strategy, but modify the components introduced in Section~\ref{sec:3} -- accordingly the ranking loss for distillation, training files for hard negatives, and/or the PLM initialization. We thus consider the following scenarios: \texttt{SPLADE}, which corresponds to the original setting~\cite{10.1145/3404835.3463098}; \texttt{DistilMSE} which relies on Eq.~\ref{mse_loss} for training; \texttt{SelfDistil} where models are trained using Eq.~\ref{eq_self};
    \texttt{EnsembleDistil} which rather makes use of Eq.~\ref{eq_ensemble}; and finally \texttt{CoCondenser-SelfDistil} and
    \texttt{CoCondenser-EnsembleDistil} which correspond to the two latter scenarios, where the model has additionally been initialized from a pre-trained CoCondenser checkpoint\footnote{Available via huggingface: \url{https://huggingface.co/Luyu/co-condenser-marco}}.

For each scenario, we train $5$ models, each configuration corresponding to different values of the regularization magnitude ($\lambda_q$, $\lambda_d$) in e.g. Eq.~\ref{splade_loss}, thus providing various trade-offs between effectiveness and efficiency -- the higher the $\lambda$s, the sparser the representations. Note that, as each procedure either modifies the loss or the input pairs, the range taken by loss values slightly differs. We therefore need to adapt the $\lambda$s for each scenario; we simply rely on grid-search, and kept five configurations which covered a broad range of effective and efficient models\footnote{Code for training and evaluating models is available at: \url{https://github.com/naver/splade}}\footnote{Checkpoints for models can be found on Hugging Face: \url{https://huggingface.co/naver}}. 
To assess model robustness to random training perturbations, we train and evaluate each configuration \emph{three times} -- corresponding to different random \emph{seeds}. We thus provide average measures alongside standard deviations.

\paragraph{\bf Evaluation}
We evaluate models on the MS MARCO development set which contains $6980$ queries, as well as the TREC DL 2019 set which provides fine-grained annotations from human assessors for a set of $43$ queries~\cite{craswell2020overview}. We report R@1k for both datasets, as well as MRR@10 and nDCG@10 for MS MARCO dev set and TREC DL 2019 respectively.
We additionally assess the zero-shot performance of our models on the BEIR benchmark~\cite{beir_2021}. For comparison with other approaches, we rely on the subset of $13$ datasets that are readily available, thus we do not consider \texttt{CQADupstack}, \texttt{BioASQ}, \texttt{Signal-1M}, \texttt{TREC-NEWS} and \texttt{Robust04}; otherwise, we evaluate our models on the complete benchmark ($18$ datasets).

As SPLADE models allow for various trade-offs -- depending on the regularization strength -- we resort to the FLOPS measure used in \cite{10.1145/3404835.3463098} to compare efficiency between models, which gives an estimation of the average number of floating point operations needed to compute the score between a query and a document, empirically estimated from the collection and a set of $100k$ development queries after training. Note that more informative metrics (e.g. query latency) could have been used. However, such measures can be hard to properly evaluate, depending on the systems; as we only compare efficiency of SPLADE models, the FLOPS is therefore informative enough.


\subsection{Results and discussion}

We compare our models to two types of approaches -- reporting results from corresponding papers:
\begin{inlinelist}    
    \item \texttt{Simple training} relying on a single round of training with BM25 negatives; it includes models like DeepImpact~\cite{10.1145/3404835.3463030}, ColBERT~\cite{colbert} or SPLADE~\cite{10.1145/3404835.3463098}
    \item \texttt{Training ++} with various training strategies to improve performance; it includes models like ANCE~\cite{xiong2021approximate}, TAS-B~\cite{Hofstaetter2021_tasb_dense_retrieval} or ColBERT v2~\cite{santhanam2021colbertv2}. We also include results from Contriever~\cite{izacard2021contriever} in the zero-shot setting (see Table~\ref{table_beir}).
\end{inlinelist}

\begin{figure}[b]
  \caption{MRR@10 (MS MARCO dev) vs FLOPS for all our scenarios. Each point corresponds to the average -- with standard deviation -- over three runs with different random seeds. Highlighted points correspond to the models for which results are reported in Table~\ref{table_1} (i.e. with FLOPS$\leq 3$).}
  \includegraphics[width=0.5\textwidth]{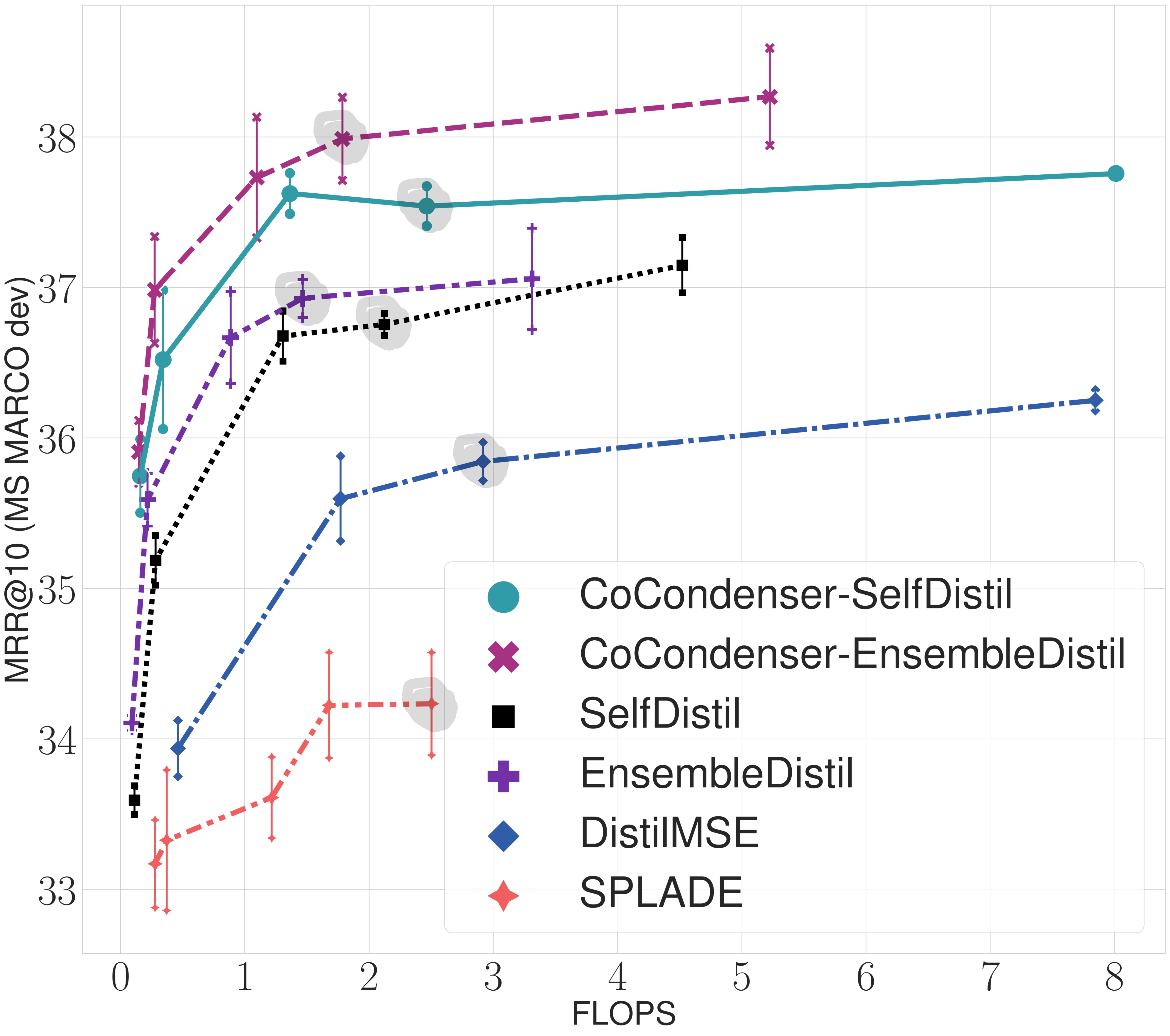}
  \label{perf_flops}
\end{figure}

\begin{figure}[b]
  \caption{Mean nDCG@10 over the 18 BEIR datasets vs FLOPS. Each point corresponds to the average -- with standard deviation -- over three runs with different random seeds.}
  \includegraphics[width=0.5\textwidth]{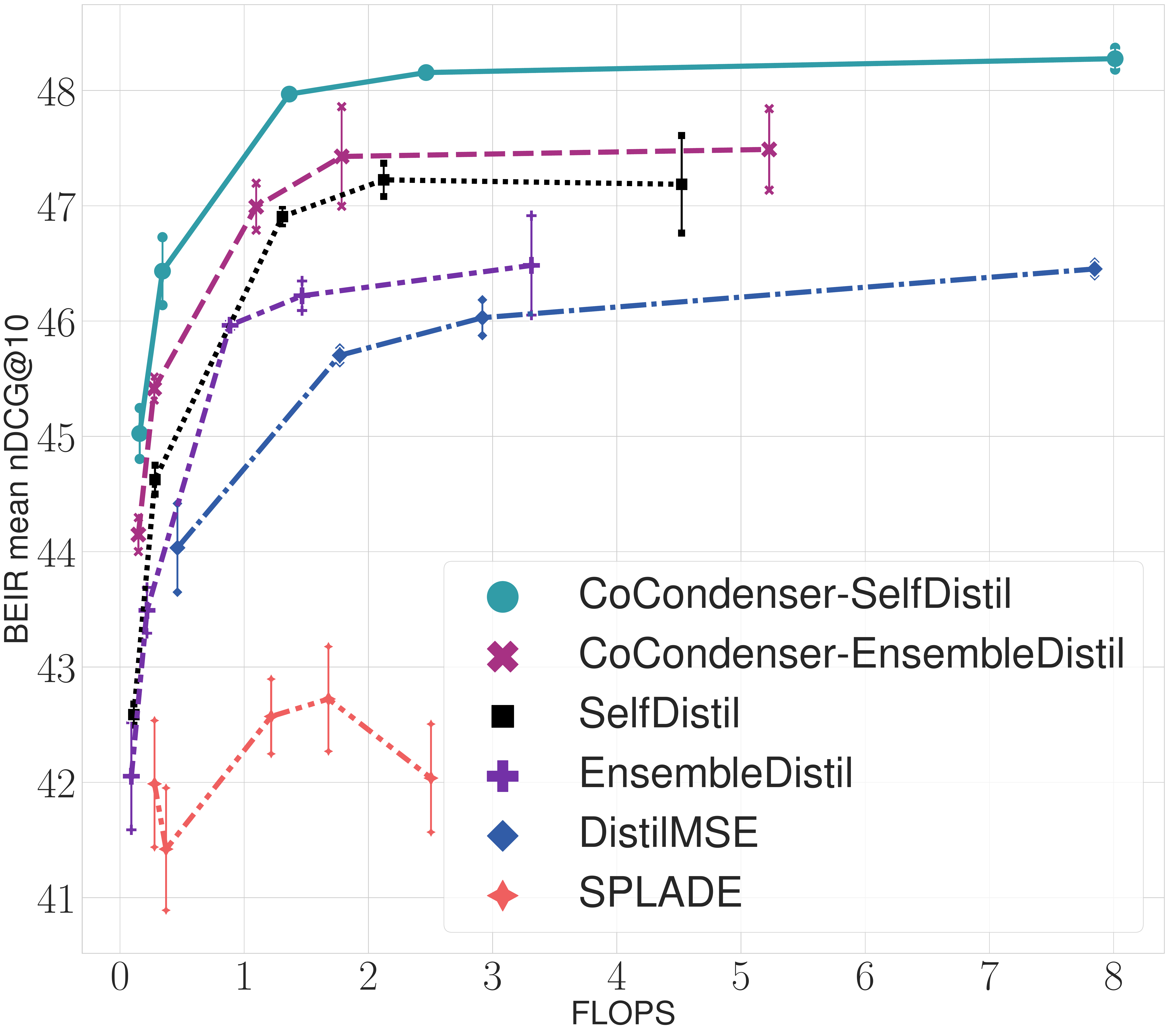}
  \label{perf_beir}
\end{figure}

MS MARCO dev and TREC DL 2019 results are given in Table \ref{table_1}. For each, we report the best model configuration (among the five that are trained), \emph{with a FLOPS value inferior to $3$}. It thus does not necessarily correspond to the best performance we can obtain, but to a more realistic trade-off between effectiveness and efficiency (i.e. model compression which depends on $\lambda$). The interplay between the two is given in Fig.~\ref{perf_flops} and Fig.~\ref{perf_beir}: we respectively report MRR@10 on MS MARCO dev and mean nDCG@10 on BEIR datasets \emph{vs} FLOPS, for the five configurations in each scenario. Please note that for BEIR, averaging metrics over multiple datasets is questionable~\cite{10.1145/3269206.3271719}: we thus provide evaluation on every dataset in Table~\ref{full_table_beir} in the Appendix section.


Overall, we observe that:
\begin{enumerate*}
    \item SPLADE is able to take advantage of various training strategies to increase its effectiveness; 
    \item Performance boosts are additive;
    \item Our scenarios lead to competitive and state-of-the-art results on respectively in-domain and zero-shot evaluation, e.g. our \texttt{CoCondenser-EnsembleDistil} reaches $38$ MRR@10 on MS MARCO;
    \item Model effectiveness is linked to efficiency (the sparser, the less effective, which is also true when evaluating on zero-shot).
\end{enumerate*}

\setlength{\tabcolsep}{2pt} 
\begin{table}
\centering
\caption{Evaluation on MS MARCO passage retrieval (dev set) and TREC DL 2019. As each scenario includes 5 models with different regularization strength, we only report the best performance with a FLOPS value inferior to 3 (see Figure~\ref{perf_flops}). As each run is trained with 3 different random seeds, we also report average measures.}
\begin{tabular}{lcccc}
\toprule
\texttt{model} &  \multicolumn{2}{c}{MS MARCO dev} & \multicolumn{2}{c}{TREC DL 2019} \\
& MRR@10 & R@1k & nDCG@10 & R@1k  \\
\midrule
\multicolumn{5}{l}{\texttt{Simple training}}  \\
\hline
BM25 & 18.4 & 85.3 & 50.6 & 74.5  \\ 
doc2query-T5~\cite{doct5} & 27.7 & 94.7 & 64.2 & 82.7\\
DeepImpact~\cite{10.1145/3404835.3463030}                         &    32.6 & 94.8 & 69.5 & - \\
SPLADE~\cite{10.1145/3404835.3463098}  & 32.2   & 95.5  & 66.5 & 81.3   \\
COIL-full~\cite{gao-etal-2021-coil}                         &     35.5 &  96.3 &   70.4 & -   \\
ColBERT~\cite{colbert} & 36.8 & 96.9 & - & - \\
\hline
\multicolumn{5}{l}{\texttt{Distillation, negative mining or pre-training}} \\
\hline
ANCE~\cite{xiong2021approximate} & 33.0 & 95.9 & 64.8 & -   \\
TCT-ColBERT~\cite{lin-etal-2021-batch} & 35.9 & 97.0 & 71.9 & 76.0  \\
TAS-B~\cite{Hofstaetter2021_tasb_dense_retrieval}       &  34.7 &  97.8 & 71.7 & 84.3  \\
RocketQA-v2~\cite{ren-etal-2021-rocketqav2} &  38.8 &  98.1 & - & -  \\
CoCondenser~\cite{DBLP:journals/corr/abs-2108-05540} &  38.2 & 98.4 & - & -  \\
AR2~\cite{https://doi.org/10.48550/arxiv.2110.03611} &  39.5 & 98.6 & - & -  \\
ColBERTv2~\cite{santhanam2021colbertv2} &  39.7 &  98.4 & - & - \\

\hline 
\texttt{Our methods: SPLADE++} & & & &  \\
SPLADE (simple training) & 34.2 & 96.6 & 69.9 & 81.5  \\
DistilMSE & 35.8 & 97.8 & 72.9 & 85.9  \\
SelfDistil & 36.8 & 98.0 & 72.3 & 86.9 \\
EnsembleDistil & 36.9 & 97.9 & 72.1 & 86.5 \\
CoCondenser-SelfDistil$^\dagger$  & 37.5 & 98.4 & 73.0 & 87.8 \\
CoCondenser-EnsembleDistil$^\ddagger$  & 38.0 & 98.2 & 73.2 & 87.5 \\
\hline 


\end{tabular}
\label{table_1}
\end{table}

\setlength{\tabcolsep}{2pt} 
\begin{table}
\small
\centering
\caption{Mean nDCG@10 on the subset of 13 BEIR datasets. SPLADE++$^{\ddagger,\dagger}$ respectively correspond to our best scenarios CoCondenser-EnsembleDistil$^\ddagger$ and CoCondenser-SelfDistil.$^\dagger$}
\begin{tabular}{lcccccc}

\toprule
 &  BM25 & TAS-B & Contriever & ColBERTv2 & \texttt{SPLADE++}$^\ddagger$ & \texttt{SPLADE++}$^\dagger$\\
\midrule
\texttt{nDCG@10} &  43.7 & 43.7 & 47.5 & 49.7 & 50.5 & 50.7 \\

\end{tabular}

\label{table_beir}
\end{table}

\paragraph{\bf Results analysis}

From Table~\ref{table_1}, we observe that our SPLADE model with \texttt{max} pooling (Eq.~\ref{eq_splad_max}) already achieves high performance, compared to models in the \texttt{simple training} case. The use of distillation (\texttt{DistilMSE} scenario) offers the largest boost in effectiveness across all scenarios ($+1.6$ MRR@10). When combining distillation with hard negative mining, we note that the \texttt{SelfDistil} case seems to be the most effective. However, when changing the initialization checkpoint to CoCondenser, we observe the reverse trend, where \texttt{CoCondenser-EnsembleDistil} is able to outperform its counterpart. One plausible explanation might be that this checkpoint builds on the \texttt{bert-base-uncased} model -- contrary to our default \texttt{distilbert-base-uncased}, the former containing almost twice as much parameters, allowing to better take advantage of various sources of negative mining. We also note that improvements are less clear when inspecting other metrics like R@1k or nDCG@10 on TREC DL 2019.

In Table~\ref{table_beir}, we report the BEIR results for our two CoCondenser scenarios, reaching state-of-the-art results on zero-shot evaluation, strengthening the observation that sparse retrieval models seem to be better able to generalize~\cite{10.1145/3404835.3463093,beir_2021,formal2021match}. We also report in Table~\ref{table_bm25} the results of the two previous approaches combined with BM25 (sum); additional gains can be obtained, showing that pure lexical approaches are still somehow complementary to sparse neural models, especially in a zero-shot setting.

\begin{table}
\begin{tabular}{lcc}

\toprule
 &  \texttt{SPLADE++}$^\ddagger$ + BM25 & \texttt{SPLADE++}$^\dagger$ + BM25 \\
\midrule
\texttt{nDCG@10} &  52.1 & 52.1  \\

\end{tabular}
\caption{Mean nDCG@10 on the subset of 13 BEIR datasets for a simple combination (sum) of \texttt{SPLADE++} and BM25.}
\label{table_bm25}
\end{table}

In Fig.~\ref{perf_flops}, we analyse the interplay between effectiveness and efficiency (in terms of FLOPS) on MS MARCO dev set. There is an overall trend that more expressive models tend to be more effective. We also observe additivity in improvements, with the best model on MS MARCO (\texttt{CoCondenser-EnsembleDistil}) taking advantage of distillation, ensemble mining and pre-trained checkpoint altogether. We observe the same kind of behavior on the BEIR zero-shot evaluation, where effectiveness comes with a higher complexity (Fig.~\ref{perf_beir}). However, the \texttt{SelfDistil} scenarios seem to be better suited for generalization (they both respectively outperform their \texttt{EnsembleDistil} counterparts, also see Table~\ref{table_beir}). One reason for this behavior might be the use of only dense retrievers in the latter case: as such models have been shown to overfit on MS MARCO, the mined negatives might be too specialized on this dataset, thus hurting generalization capabilities of subsequently trained models. Finally, we see that overall, results are rather stable under varying random seeds.

\section{Conclusion}
In this paper, we have built on the SPLADE model, and studied to which extent it is able to take advantage of training improvements like distillation and hard negative mining, as well as better suited PLM initialization: combined altogether, the resulting model reaches state-of-the-art performance on both in-domain and zero-shot evaluation. We also investigated the link between effectiveness and efficiency -- induced by the degree of regularization -- highlighting that more expressive models are better at generalization.



\bibliographystyle{ACM-Reference-Format}
\balance
\bibliography{sample-base}

\appendix

\section{Detailed BEIR evaluation}

We provide in Table~\ref{full_table_beir} the complete evaluation on BEIR datasets. Overall, SPLADE variants obtain results that are state-of-the-art.

\setlength{\tabcolsep}{6pt}
\begin{table*}[b]
\caption{ndcg@10 on BEIR for all the datasets (18). For comparison, we report results directly from corresponding papers, where the evaluation is generally done on the subset of 13 readily available BEIR datasets.}
\label{full_table_beir}
\begin{tabular}{@{}lcccccc@{}}
\toprule
\textbf{Corpus} & \textbf{BM25} & \textbf{TAS-B} & \textbf{Contriever} &\textbf{ColBERTv2} & \textbf{SPLADE++}$^\ddagger$ & \textbf{SPLADE++}$^\dagger$ \\
\toprule
\texttt{TREC-COVID}                 &    65.6      &     48.1     &   59.6   & \textbf{73.8}    & 72.7  & 72.5  \\
\texttt{BioASQ}                 & 46.5          &   38.3       &   -        & - &  49.7 & \textbf{50.8} \\
\texttt{NFCorpus}                 & 32.5           &     31.9     &     32.8    & 33.8  &  \textbf{34.8} & 34.5 \\
\texttt{NQ}                 &   32.9        &     46.3     &   49.8        & \textbf{56.2} &  53.7 & 53.3 \\
\texttt{HotpotQA}                 & 60.3           &   58.4       &  63.8  &   66.7    & 68.7 & \textbf{69.3} \\
\texttt{FiQA-2018}                 & 23.6           &       30.0   &   32.9      & \textbf{35.6} & 34.8 &  34.9 \\
\texttt{Signal-1M (RT)}                 &     \textbf{33.0}      &  28.9        &  - &    -    &  30.0 &  30.9 \\
\texttt{TREC-NEWS}                 & 39.8           &  37.7        &  -  &    -   & 41.5 & \textbf{41.9} \\
\texttt{Robust04}                 & 40.8           &    42.7      &   -   &   -  &  46.7 &  \textbf{48.5} \\
\texttt{ArguAna}                 & 31.5           &   42.9        &  44.6  &  46.3  & \textbf{52.5}    & 51.8 \\
\texttt{Touché-2020}                 &  \textbf{36.7}         &    16.2      &  23.0   & 26.3      & 24.5 & 24.2 \\
\texttt{CQADupStack}                 &  29.9         &   31.4       &   34.5    &  -  &  33.4 & \textbf{35.4} \\
\texttt{Quora}                 &  78.9         &   83.5       &   \textbf{86.5}        & 85.2  & 83.4 & 84.9 \\
\texttt{DBPedia}                 & 31.3          &       38.4   &     41.3      & \textbf{44.6} & 43.6 & 43.6 \\
\texttt{SCIDOCS}                 & 15.8          &      14.9    &    \textbf{16.5}       & 15.4 & 15.9 & 16.1 \\
\texttt{FEVER}                 &  75.3         &    70.0      &    75.8       & 78.5  & 79.3 & \textbf{79.6} \\
\texttt{Climate-FEVER}                 &   21.3        &  22.8        &    23.7       & 17.6 & 23.0 & \textbf{23.7}  \\
\texttt{SciFact}                 & 66.5          &   64.3       &   67.7        & 69.3 & 70.2 & \textbf{71.0} \\ \bottomrule
\end{tabular}
\end{table*}

\end{document}